# Role of Delay in Brain Dynamics

Yuval Meir[a], Ofek Tevet[a], Yarden Tzach[a], Shiri Hodassman[a] and Ido Kanter[a,b,*]

[a]Department of Physics, Bar-Ilan University, Ramat-Gan, 52900, Israel.
[b] Gonda Interdisciplinary Brain Research Center, Bar-Ilan University, Ramat-Gan, 52900, Israel.

* Corresponding author at: Department of Physics, Bar-Ilan University, Ramat-Gan, 52900, Israel. E-mail address: ido.kanter@biu.ac.il (I. Kanter).

## Abstract

Significant variations of delays among connecting neurons cause an inevitable disadvantage of asynchronous brain dynamics compared to synchronous deep learning. However, this study demonstrates that this disadvantage can be converted into a computational advantage using a network with a single output and $M$ multiple delays between successive layers, thereby generating a polynomial time-series outputs with $M$. The proposed role of delay in brain dynamics (RoDiB) model, is capable of learning increasing number of classified labels using a fixed architecture, and overcomes the inflexibility of the brain to update the learning architecture using additional neurons and connections. Moreover, the achievable accuracies of the RoDiB system are comparable with those of its counterpart tunable single delay architectures with $M$ outputs. Further, the accuracies are significantly enhanced when the number of output labels exceeds its fully connected input size. The results are mainly obtained using simulations of VGG-6 on CIFAR datasets and also include multiple label inputs. However, currently only a small fraction of the abundant number of RoDiB outputs is utilized, thereby suggesting its potential for advanced computational power yet to be discovered.

## 1. Introduction

The dynamics of the brain differ from those of complex electronic circuits such as computers. One of the remarkable differences is that the dynamics of electronic devices are synchronous and controlled by the central clock of the device (e.g., the CPU clock), whereas brain dynamics are asynchronous. This difference stems from the typically different ratios between the fundamental timescales that control their dynamics.

The synchronous dynamics of electronic circuits are based on the following components. The rise time, $t_{RT}$, of a basic electronic element, such as a transistor, can be in the order of $\sim 10^{-11}$ s for a 100 $GHz$ clock speed. However, the necessary ingredient to achieve synchronous dynamics is the fluctuations in $t_{RT}$, which are typically expected to be two orders of magnitude below, that is, $\Delta(t_{RT}) \sim 10^{-13}$. This requirement is necessary to reduce the accumulation of unpredicted delays in processing pipeline operations on a chip.

The density of transistors in advanced chips can exceed $O(10^8)$ per millimeter square [1, 2], indicating $\sim 0.1\ \mu m$ distance between two neighboring transistors. Because the velocity of the electronic signal is in the order of the speed of light, that is, $\sim 10^8\ m/s$, the time-lag for the propagation signal between two neighboring transistors is $t_{prop} = O(10^{-15}\ s)$ with negligible relative fluctuations. Thus, in electronic devices, the following order of timescales is obtained:

$$t_{prop} < \Delta(t_{RT}) < t_{RT} \quad (1)$$

where the rightmost inequality is the necessary condition for the formation of synchronous dynamics.

In brain dynamics, the distance of an unmyelinated axon is typically in the order of millimeters, and the speed of the signal is approximately a meter per second thus, $t_{prop}$ is of the order of several milliseconds (ms) [3]. Similarly, the rise time of a neuron, $t_{RT}$, including the propagation of the signal via its dendrites, measured by the neuronal response latency [4, 5], is several ms too. Finally, the fluctuations in $t_{prop}\ and\ t_{RT}$, depending on the recent historical activity of the neurons, are also in the order of several ms, where an exceptional case is neurons stimulated at fixed and low frequencies [5]. Thus, in terms of brain dynamics,

$$O(t_{prop}) = O\left(\Delta(t_{prop})\right) = O(\Delta(t_{RT})) = O(t_{RT}) = O(ms) \quad (2)$$

Under such timescales, the implementation of synchronous dynamics is impossible and asynchronous brain dynamics are inevitable.

The implementation of reliable computations in the brain under such poor reality (Eq. (2)) requires two components. The first is the temporal summation of the neuronal input signals from many connecting neurons. As all these incoming signals are asynchronous, an integration process by the membrane potential over a time scale of approximately $15\ ms$ is required. In computational modeling, this is known as a leaky and integrate fire neuron [6]. The second component is population dynamics [7], wherein a perceptual entity is represented by many neurons, for example, 100. This component overcomes neuronal response failures and enhances the signal-to-noise ratio, where the output signal is distributed over time (Eq. (2)). The integrate and fire neurons and population dynamics substantially slow down the dynamics and waste biological hardware, neurons, and synapses.

Asynchronous brain dynamics are an inevitable disadvantage stemming from the poor biological hardware. To overcome these poor dynamics, additional biological hardware and a considerably longer computational time are required. The question at the core of this research is whether this conclusion is correct, or whether brain dynamics conceals certain advanced computational features that are absent in commonly used artificial synchronous dynamics.

Mimicking the dynamics of the entire brain does not provide the opportunity to independently examine each of its time-dependent components (Eq. (2)). To achieve this goal, this study adopts a perturbation around standard deep learning [8] and approaches the brain dynamics by relaxing one of its constraints: multiple delays. Specifically, several delays are allowed between consecutive layers of the deep architecture. The results indicate that variation in delays is not an inevitable disadvantage of biological systems; rather, they facilitated a new computational hypothesis that offers several advantages in brain dynamics. This new hypothesis, the role of delay in brain (RoDiB), is presented and exemplified on a classification task, CIFAR-M [9-12], using VGG-6 deep architectures [11, 13].

## 2. Results
### 2.1. RoDiB model

A typical deep architecture for a classification task comprises three components. An input image, a series of CLs, and an FC layer to the $M$ output units representing the labels (Fig. 1A). Each CLs comprised a set of filters (Fig. 1A, gray) terminates at a layer of output units, optionally followed by a pooling operator, batch normalization, and ReLU activation functions (Fig. 1A, light blue) [14]. An example of such architecture is VGG-6 [11, 15] trained on CIFAR-M.

An alternative architecture for solving the same classification task is a set of $M$ replicated architectures [16], each with independent weights and a single output representing a specific label (Fig. 1B). The activation function of the output units is Sigmoid, and each architecture is trained to achieve an output equals to one for its selected label and zero for the remaining, $M-1$, labels. The training of such $M$ VGG-6 systems with a single output each on CIFAR-M indicates the same accuracies as that for a single VGG-6 system with $M$ outputs (Table I). Nevertheless, the M system solution (Fig. 1B) requires significant additional hardware, weights, and nodes compared with a single system with multiple outputs (Fig. 1A).

The third learning strategy partially overcomes the disadvantage of the second one by folding the $M$ systems (Fig. 1B) into a single system at the expense of multiple delays (Fig. 1C). Specifically, each interior CLs is connected to its successive layer using $M$ delays (Fig. 1C), where the input and the FC layers are exceptional, and each comprises a single delay. The number of units in such an architecture is slightly lower than that of a single system (Fig. 1A), minus $M-1$ for the output layer; however, the number of CL weights is approximately $M$ times. This RoDiB system generates polynomial time-series outputs with $M$, $M^4$ for VGG-6, owing to the different delay routes to the single output (Fig. 1C). Training each label on $M$ non-overlapping delay routes for the interior CLs, denoted as $[1,1,1,1]$ ... $[M,M,M,M]$ (Fig. 1D), results in the same accuracies as that for a single system (Table I). The output units are spanned in time (Fig.

1D) instead of space (Fig. 1A), with additional $M^4 - M$ outputs, which may also be useful. It is evident that additional multiple $M$ delays for the input and FC layers (Fig. 1C) results in $M$ completely non-overlapping routes (Fig. 1B); however, merging routes in the first and last layers were found to enhance accuracies (not shown).

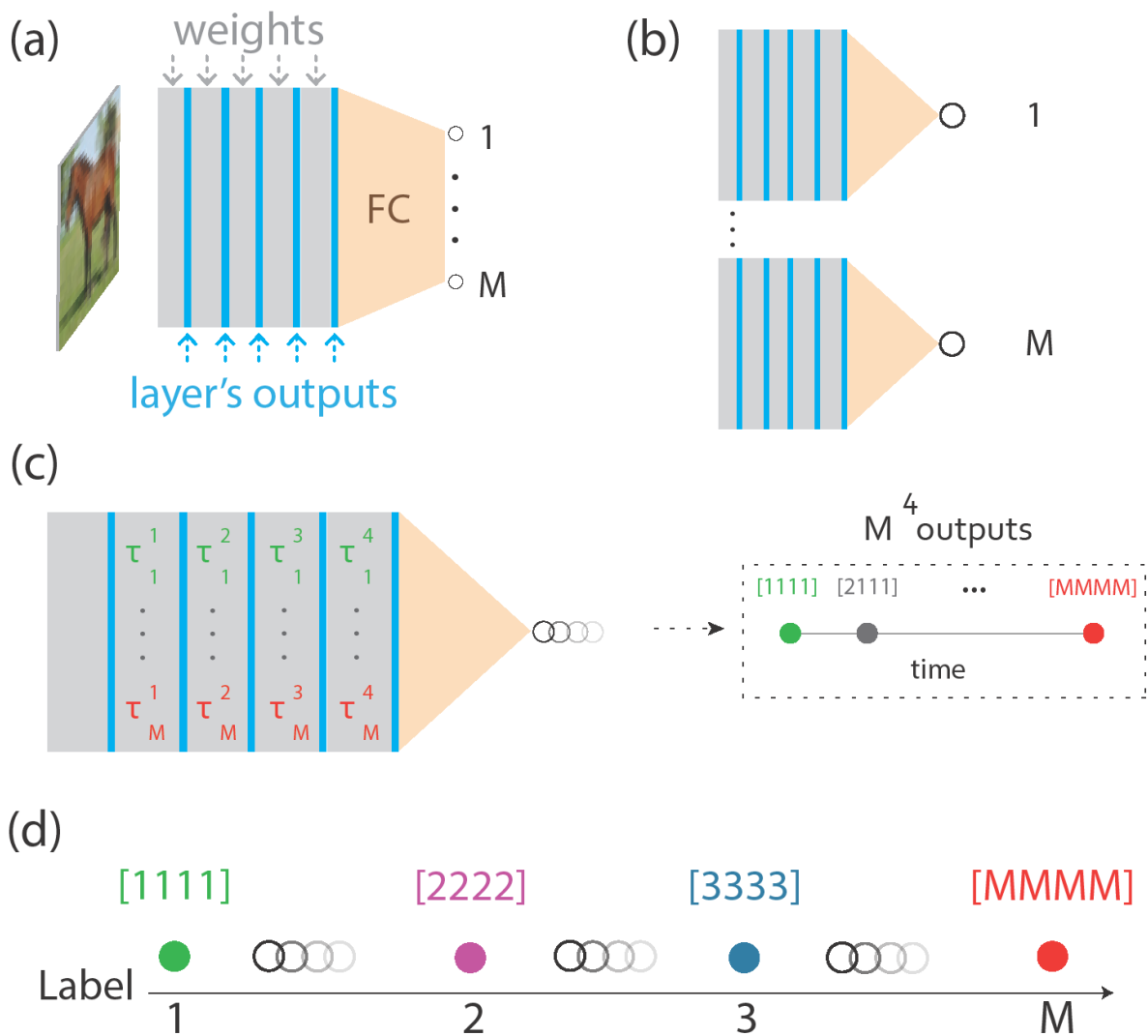

**Fig. 1.** The RoDiB model. (a) Scheme of a standard deep architecture for classification of $M$ labels comprising six layers, five CLs (gray) with an output layer (light blue), and one FC layer (orange). (b) Scheme of $M$ architectures as in panel (a), but with a single output,

representing a specific label. (c) Scheme of the RoDiB architecture, where weights for the four interior CLs comprise $M$ delays each, generating a time series of $M^4$ outputs (right). (d) Scheme of selected $M$, non-overlapping training routes (in the four interior CLs, color coded), for training the $M$ labels (the rest $M^4 - M$ routes denote by chains of faded black circles).

| | VGG-6 | | |
|---|---|---|---|
| $CIFAR - M$ | $Signle\ sys.$ | $M\ sys., single\ output$ | $RoDiB - M$ |
| 10 | 0.932 | 0.932 | 0.935 |
| 40 | 0.832 | 0.833 | 0.835 |

**Table I:** Accuracies for CIFAR-10 and CIFAR-40/100, 40 labels out of 100, measured for three different architectures: a single systems with $M$ outputs (Fig. 1A), $M$ systems with a single output each (Fig. 1B) and the RoDiB systems (Fig. 1C).

One mission of deep learning is to classify a set of labels with high accuracy. This mission is also common to the learning process in the brain; however, knowledge development frequently includes learning newly emerging labels (Fig. 2A). This dynamic scenario in deep learning can be easily implemented by increasing the number of output units and the size of the FC layer (Fig. 2A). Next, the extended network is retrained using transfer learning, which represents a perturbation around the existing learning with fewer labels. However, the brain's hardware is less flexible; the number of neurons is practically fixed, and changing the learning architecture or strategy is not a simple task that can be implemented efficiently and frequently. Thus, the brain requires a fixed architecture that is sufficiently flexible to learn an increasing number of labels, which can be realized by the RoDiB model (Fig. 2B). For instance, assume that such a network is constructed from $M = 40$ (Fig. 1C), which initially learns 10 labels, using 10 non-overlapping routes (Fig. 1D). The learning of additional labels can be implemented on the 30 remaining non-overlapping routes, which requires retraining of the entire system (Table I). A more attractive solution would be to set $M$ small, for example, 10, and utilize the additional time-series outputs, $M^4 - M$, to learn the additional labels (Fig. 2B). This scenario is discussed in the following section.

## 2.2. RoDiB with multi-label inputs

The partial utilization of the polynomial time-series outputs, $M^4 - M$, of the RoDiB system (Fig. 1C) is examined using an additional multi-label classification [17]. In addition to the classification of the $M$ single input labels, high accuracy is required to identify additional $K$ inputs, each of which combines a pair of input labels (Fig. 3A). An input pair is formed by joining two $32 \times 32$ CIFAR inputs compressed to $32 \times 16$ [18]. The learning accuracies of $M$ singles and $K$ pairs is examined using the following scenarios (Table II). The first scenario is a single system with $M + K$ output units (Fig. 1A), whereas the second one is $M + K$ systems with a single output each (Fig. 1B). The third learning scenario is based on the RoDiB system (Fig. 1C) using $M$ non-overlapping routes for the training of $M$ singles (Fig. 1D) and combined routes for the training of K pairs. Specifically, for an input pair $(i, j)$ the selected training route is $[iijj]$ (Fig. 3B), combining the routes $[iiii]$ and $[jjjj]$ of the two single inputs (Fig. 1D). The fourth scenario is also based on the RoDiB system; however, random routes, among $M^4 - M$, are selected for training the pairs (Fig. 3C). Limited results indicate that the RoDiB system with combined routes achieves the same accuracy as a standard single system with $M + K$ outputs, which is slightly higher than RoDiB system with random routes (Table II). Hence, the RoDiB system with a flexible number of classified inputs can compete with standard deep learning strategy.

(a)

| $(Singles, Pairs)$ | VGG-6, CIFAR-10 | | | |
|---|---|---|---|---|
| | $Single\ sys.$ | $RoDiB - M + K$ $Different\ Routes$ | $RoDiB - M$ $Combined\ Routes$ | $RoDiB - M$ $Random\ Routes$ |
| (10,5) | 0.934 | 0.936 | 0.936 | 0.932 |
| (10,10) | 0.913 | 0.910 | 0.913 | 0.891 |
| (10,20) | 0.870 | 0.868 | 0.866 | 0.825 |
| VGG-6, CIFAR-20 | | | | |
| (20,10) | 0.911 | 0.908 | 0.907 | 0.890 |

(b)

| $CIFAR - M$ (Singles, Pairs) | Signle sys. | $RoDiB - M$ |
|---|---|---|
| (10,0) | 0.586 | 0.679 |
| (20,0) | 0.565 | 0.683 |
| (40,0) | 0.385 | 0.545 |
| (10,10) | 0.460 | 0.570 |

VGG-6, $d = 2$

**Table II:** (a) Accuracies of the RoDiB system trained on VGG-6 with CIFAR-10 and CIFAR-20/100 using $Singles + Pairs$ input labels. Accuracies are measured for the following three different scenarios for the trained routes of the pairs: different routes indicating that $M = Singles + Pairs$; combined routes (Fig. 3B); and random routes (Fig. 3C). (b) Similar to panel (a) for VGG-6 with $M = 10, 20$, and $40$ and $d = 2$ filters in the first CL, resulting in 16 filters in the last CL.

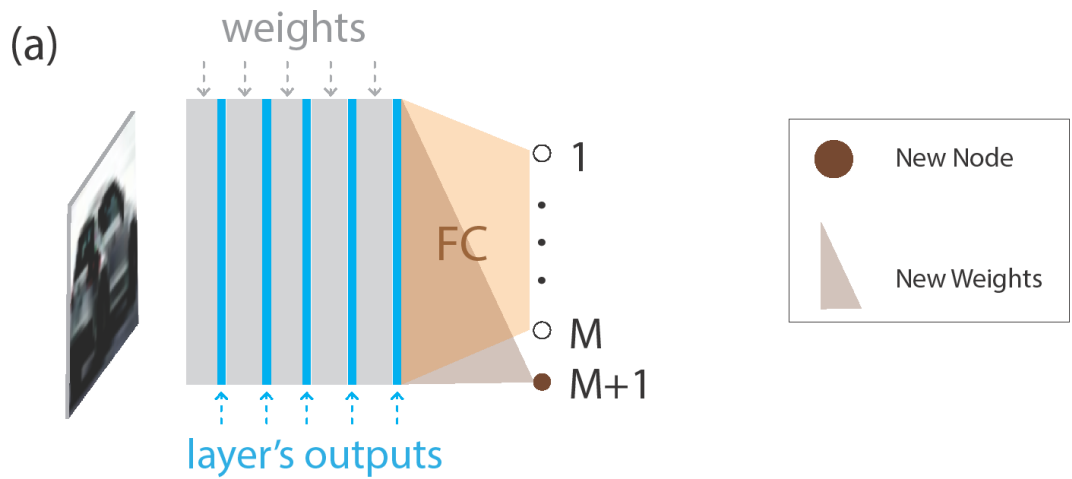

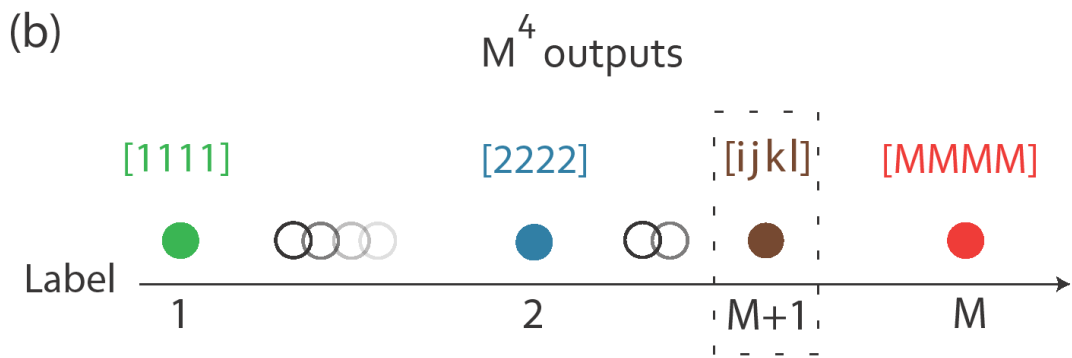

**Fig. 2.** Flexibility of the RoDiB model for brain dynamics. (a) Scheme of the architecture (Fig. 1A) and the required new hardware, a node (brown) and FC weights (triangle gray), for learning the $M+1$ new label. (b) RoDiB solution, where the new label, $M+1$, is selected as an output $[ijkl]$, among the currently $M^4 - M$ unselected time series outputs.

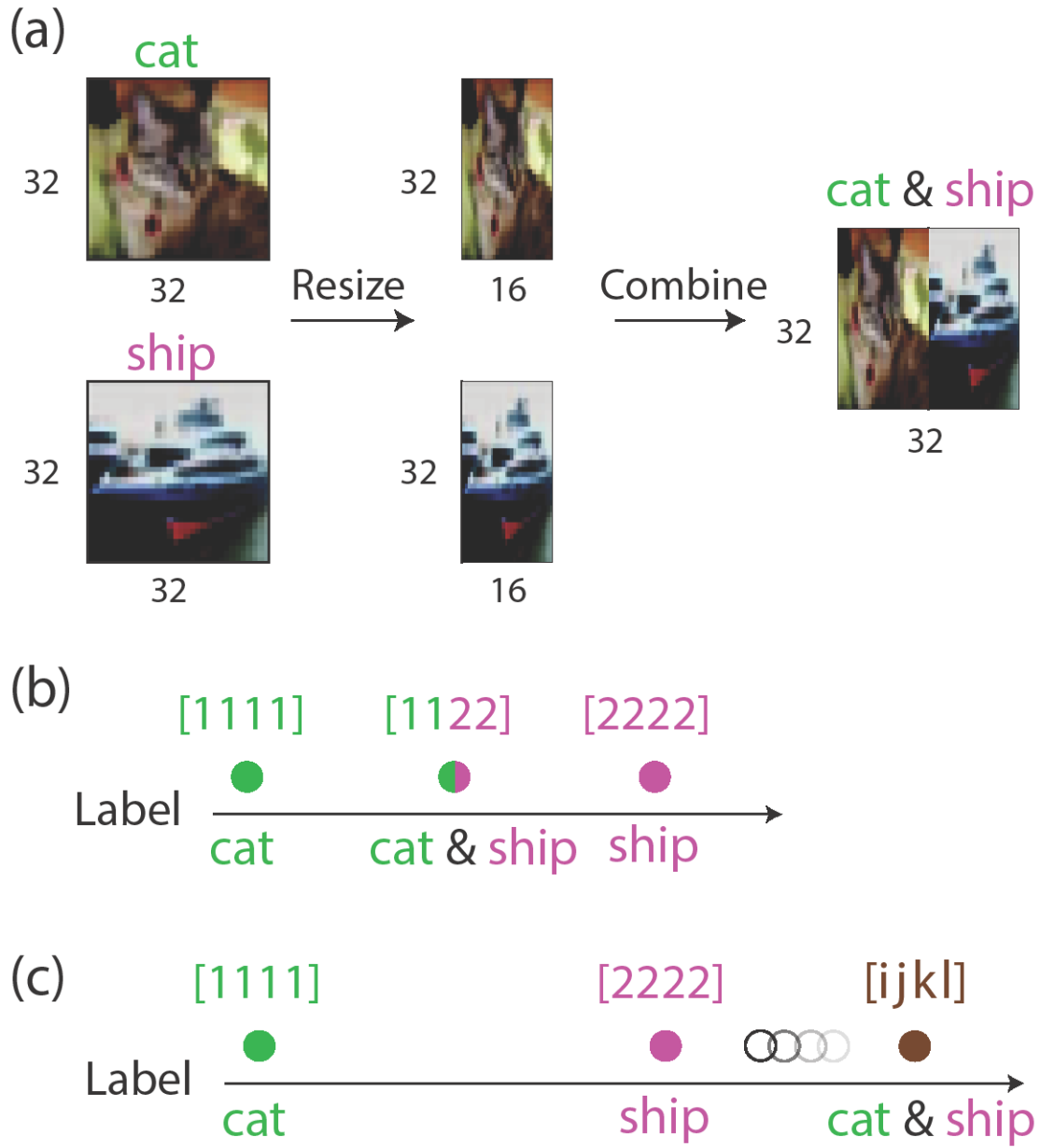

**Fig. 3.** RoDiB system with additional multi-label inputs. (a) Example of two $32 \times 32$ CIFAR-10 inputs compressed to $32 \times 16$ (left) and their subsequent merging into a $32 \times 32$ pair (right). (b) Scheme of selected non-overlapping training routes for single label inputs 1 and 2, and a combined training route [1122] for their input pair. (c) Same as in panel (b); however, a random route is selected for the training of their input pair.

*2.3. RoDiB model versus deep learning*

The presented results indicate the same accuracy for the RoDiB system and its counterpart standard deep learning architecture, with an output unit for each label (Fig. 1). Nevertheless, enhanced accuracy for the RoDiB system is expected in the limit where the number of labels $M$, is greater than the number of input units to the FC layer $N_{FC}$. For such a limit

$$M > N_{FC} \quad (3)$$

where the FC input dimension is smaller than the output dimension, the accuracy is expected to deteriorate for standard architectures, however, for the single-output RoDiB system, a relatively enhanced accuracy is expected. The crossing of this limit (Eq. (3)) is expected as the system is required to classify an increasing number of labels.

Examination of this hypothesis through simulations is difficult for the following several reasons. For a standard trained VGG-6 architecture, $N_{FC} = 512$ and the number of labels in a dataset is typically considerably smaller. Further, an effort to significantly increase $M$ and optimize its RoDiB system accuracy (Fig. 1C) results for us in unattainable computational complexity. Thus, we considered another limit of VGG-6, where the number of filters in the first layer is $d = 2$, instead of 64; accordingly, $N_{FC} = 16$ ( as $512/64 = 16/2$). Simulations around this limit (Eq. (3)) of VGG-6 with $d = 2$ filters in the first CL indicate that the optimized accuracies of the RoDiB system are significantly enhanced compared with the standard architecture (Table III).

## 3. Discussion

The brain dynamics are asynchronous, whereas the dynamics of artificial learning and electronic chips are synchronous. Synchronous dynamics do not necessarily indicate that all the elements of the integrated system respond simultaneously; however, their relative activation time-lags remain fixed throughout the dynamics. This is a necessary condition for maintaining the correct pipeline of operations and ensuring a reliable global role of the system. Inherent asynchronous brain dynamics are primarily attributable to significant variations in delays among connecting neurons and delays in their internal dynamics, neuronal response latencies. Asynchrony is reinforced by significant time-dependent variations in these delays as a function of recent activity of the network, including neuronal response failures. The use of imprecise and unreliable biological hardware inevitably leads to enhanced asynchronous dynamics, whereas ultrafast electronic devices are precise and reliable. A common manner through which the brain overcomes these limitations and maintains reliable operations involves the use of integrate and fire neurons and population dynamics. However, these ingredients significantly slow down the dynamics and result in low hardware efficiency.

In contrast, this study found the advantages of brain asynchronous dynamics, which are absent from artificial synchronous dynamics. Examination at the network level is required to isolate variations among delays from all other features of brain dynamics, including internal neuronal dynamics and population dynamics. Thus, the realization of classification tasks using standard deep learning architectures was investigated under the perturbation of multiple delays between consecutive CLs and with a single output unit (Fig. 1). For a particular input, this RoDiB system generated a time-series of outputs, which enabled the spanning of the probability for each predicted label in time (Fig. 1C) instead of in space (Fig. 1A). Spanning outputs in space or time resulted in similar accuracies, as demonstrated in the limited simulations for the VGG-6 and CIFAR datasets.

A byproduct of multiple delays connecting successive CLs is an extended time-series of outputs. At the expense of $M$ delays between successive CLs, a polynomial number of outputs with $M$ is generated, whose power increases linearly with the number of CLs. The additive effect on the number of convolutional weights, excluding the decrease in the FC layer (Fig. 2), results in a multiplicative effect on the number of outputs. These enormous outputs can classify additional labels, at least with the same accuracy as standard deep architectures (Table II). Nevertheless, the advantage of the RoDiB system is its flexibility in classifying emerging new labels without expanding the architecture to include more neurons and weights (Fig. 2), which are impractical brain processes. The extension of the RoDiB model to include hierarchical timescales among the delays may better describe brain dynamics [19, 20].

Another advantage of the RoDiB system is its enhanced accuracy compared with standard deep architectures (Fig. 1A). This enhancement is evident in the limit when the number of input units in the FC layer is less than the number of output labels (Table II). Mapping from a low-dimensional input to a higher one results in decreased accuracy. However, for the RoDiB system, this limit is unachievable because it comprised a single output, independent of the number of labels. This limit emerges for a dynamically increasing number of required classified labels as in life realities, or for architectures with a small number of filters per CL (Table II).

The RoDiB model is exemplified on an artificial convolutional neural network, VGG-6, which is far from brain architecture. Preliminary results indicate that adding delays to a dendritic tree enhance accuracies [21]. Specifically, our previous work indicated that a 3-layer tree architecture inspired by experimental-based dendritic tree adaptations outperforms a LeNet-5 feedforward network on classification of CIFAR-10 dataset [21]. Adding ten delays to the intermediate layer of the tree, the sampling layer, enhances accuracies in comparison to both Tree-3 architecture without delays and LeNet-5.

In addition, the RoDiB model considers only one aspect of the delays, which is the generation of enormous time-series outputs, and neglects possible interference among different delayed routes to the output. In a feed-forward architecture composed of leaky integrate and fire neurons, an input is presented several times at a high frequency to the network, and the relative firing of the output units represents the label likelihoods. Recent in-vitro and in-vivo experiments have indicated a long-term neuronal plasticity mechanism that temporarily silences neurons following their recent spiking activity [22]. This silencing mechanism suggests a reliable method for sequence identification without the need for recurrent neural networks, where different input labels are trained on different sub-networks. For the RoDiB model, even within a single input frame, a neuron is stimulated several times, in the range $M$ to $M^4$ in the FC and CLs (Fig. 1C), and can generate silencing periods. Hence, different delay routes to the output (excluding the shortest routes) act on different dynamically created feedforward sub-networks. Specifically, the outputs generated from the latest input frames of each input label are expected to be stochastically trained on different subnetworks. Training based only on the latest input frames of each input may result in enhanced accuracy. Preliminary results indicate that a RoDiB system with and without a specific implementation of a silencing mechanism achieve similar accuracy. The possibility of enhanced accuracy using a version of the neuronal silencing mechanism warrants further investigation.

Further research must extend the presented results of the RoDiB model to other datasets and deep architectures and consider several directions, including the following: First, it is possible for frequent input labels or important labels to be trained on routes with short delays, implementing a short latency between the presented input and its identification (Fig. 1C). In principle, one cannot identify the maximal output before the end of the time-series; however, a probabilistic inference can be made when the gap between one of the first several outputs is large compared to the others. The second is the interplay between the correlations among the training routes of related inputs and their effects on the accuracy. For instance, if two labels are corrected, or if there are inputs combining both, it would be interesting to examine the optimal training routes for such correlated inputs. In addition, abundant polynomial delay routes can be utilized to enhance the accuracy using a committee among several training routes for the same label [23]. Third, different accuracies can be found for correlated inputs in comparison with other uncorrelated inputs; for example, multiple input labels (Fig. 3) and labels that appear only as a single input label. However, the optimal training strategy for such scenarios remains unclear. Another interesting generalization is the utilization of multiple delays using restricted Boltzmann machines [24-26]. All of these further research directions must bear in mind that the running time per epoch significantly increases with $M$.

Preliminary results indicate the useful utilization of the abundance of outputs, $M^4$, of the RoDiB system. For RoDiB VGG-6 with $d = 2$ and $M = 10$ (only $16$ outputs) the accuracy for classifying CIFAR-10 is ~$0.92$, just like VGG-6 with $d = 64$ (without delays). The number of filters in the entire RoDib system, $d = 2$, is only $462$, which is less than the number of filters in the last convolutional layer with $d = 64$, $512$. In addition, the obtained accuracy for CIFAR-40/100, RoDiB VGG-6 with $d = 3$ (only $81$ outputs) is

~0.825, similar to VGG-6 with $d = 64$ [11]. These results indicate that a large fraction of the RoDiB outputs can be used for classification without affecting accuracies.

Finally, it would be interesting to obtain a crude estimate of the order number of delays, $M$, between successive layers in brain activity. The unit time resolution is composed of a spike duration, $\sim 1\ ms$, and its consecutive absolute refractory period, $\sim 2\ ms$, $3\ ms$ in total. The delay between connecting neurons in successive layers, including interneuron delays and neuronal response latencies, is typically estimated to be of the order of a few tens of ms. Hence, the practical $M$ in brain dynamics is of the order of 10. However, it must be considered that brain activity is more complex, where, for instance, refractory periods can be much longer and time-dependent as a function of the historical activity of the neurons [27, 28], as well as the neuronal response latencies.

## 4. Materials and methods
### 4.1 Architecture and Datasets
VGG-6 [13] with batch normalization [29] was the baseline architecture for all experiments, where RoDiB has the same architecture but with several delays and routes. All architectures were trained to classify CIFAR-10 [12], and CIFAR-K/100 [30, 31] where $K$ is the selected number of trained labels. The $K$ labels are homogeneously selected over all the classes' categories in order to reduce fluctuations in the measured accuracies, i.e. the same number of subclasses from each one of the 20 super-classes.

The singles in Table II represent images with a single label while the pairs are images with two different labels. For CIFAR-10, $Pairs = 5$ indicates non-overlapping pairs, such that each label appears only in one pair, and $Pairs = 10$ indicates that each label appears twice. Similarly, $Pairs = 20$ indicates that each label appears in four pairs.

For $M$ different systems with a single output and for RoDiB system with $M$ or $M + K$ different routes, the architecture is the same as VGG-6, with only one output node, where $M$ represents the number of delays between successive convolutional layers (CLs).

For RoDiB system with random routes the architecture was VGG-6 with a single output node, where an input associated with each one of the $M$ labels is trained on $M$ preselected different random routes. Similarly, an input combining a pair of labels is trained on either a preselected random route or combined routes, as explained in the Results section.

For RoDiB architectures, the first CL and the fully connected (FC) layer were shared among all the delay routes. For all architectures the ReLU [32] activation function was applied along the layers, except for the output layer where Sigmoid was applied. Similar results were obtained using Sigmoid activation functions for the hidden units as well.

### 4.2 Data preprocessing

Each input pixel of an image ($32 \times 32$) from the CIFAR-10 and CIFAR-K/100 databases was divided by the maximal pixel value, $255$, multiplied by $2$, and subtracted by $1$, such that its range was $[-1,1]$. During the training phase, data augmentation was used, derived from the original images, by random horizontally flipping and translating up to four pixels in each direction [33, 34].

For pairs, the images were also rescaled from their initial size of ($32 \times 32$) to ($32 \times 16$) [18] and stacked together to create an image with the initial size ($32 \times 32$).

### 4.3 Optimization

The binary cross-entropy cost function [24, 35, 36], without the Softmax normalization, was selected for the classification task and was minimized using the stochastic gradient descent algorithm [37]. The number of output units was equal to the summation of the number of single labels and pairs. The maximal accuracy was determined by searching through the hyper-parameters (see below). Cross-validation was confirmed using several validation databases, each consists of a randomly selected fifth of the training set of examples. The Nesterov momentum [38] and L2 regularization method [39] were applied.

### 4.4 Hyper-parameters

The hyper-parameters $\eta$ (learning rate), $\mu$ (momentum constant [38]), and $\alpha$ (regularization L2 [39]) were optimized for offline learning, using a mini-batch size of 100 inputs (Sensitivity of results to mini-batch sizes deserves further research [40]). The learning rate decay schedule [41] was also optimized. A linear scheduler was used such that it was multiplied by the decay factor, $q$, every $\Delta t$ epochs, and is denoted below as $(q, \Delta t)$. Different hyper-parameters were selected for each one of the trained classification task and architecture.

All the networks were trained with 200 epochs and the hyper-parameters for each layer, respectively, used for Table I and Table II(a) were $\eta = [5e-3, 1.9e-2, 1.5e-2, 1e-2, 9.5e-3, 1e-3]$, $\mu = [0.975]$, $\alpha = [3.74e-3, 6e-4, 6e-4, 6e-4, 6e-4, 4e-3]$,

where the learning rate decay schedule for all CLs was

$$(q, \Delta t) = \begin{cases} (0.65, 20) & epoch < 160 \\ (0.55, 20) & epoch \geq 160 \end{cases}$$

and $(q, \Delta t) = (0.65, 20)$ for the FC layer.

For the case $(10, 0)$ in Table II(b), the hyper-parameters for *Single sys.* were $\eta = 7.5e-2$, $\mu = 0.93$, $\alpha = 1e-4$ and for RoDiB, $\eta = 3e-2$, $\mu = 0.91$, $\alpha = 5e-4$, where $(q, \Delta t) = (0.6, 20)$ for all layers in both cases.

The hyper-parameters used for $(10, 10)$ in Table II(b) trained on a *Single sys.* were the same as used for the case of $(10, 0)$, and for RoDiB they were the same as used for Table I and Table II(a).

For $(20, 0)$ in Table II(b) the hyper-parameters for *Single sys.* were $\eta = 7.5e-2$, $\mu = 0.965$, $\alpha = 3e-4$, for each layer, respectively. For RoDiB, $\eta = 7.5e-2$, $\mu = 0.91$, $\alpha = 5e-4$, and $(q, \Delta t) = (0.6, 20)$ for all layers in both cases.

For $(40, 0)$ in Table II(b) the hyper-parameters for *Single sys.* and also for RoDiB were $\eta = [5e-3, 1.9e-2, 1.5e-2, 1e-2, 9.5e-3, 1e-3]$, $\mu = [0.975]$, $\alpha = [3.74e-3, 6e-4, 6e-4, 6e-4, 6e-4, 4e-3]$, for each layer, respectively, where the learning rate decay schedule for all CL layers was

$$(q, \Delta t) = \begin{cases} (0.65, 20) & epoch < 160 \\ (0.55, 20) & epoch \geq 160 \end{cases}$$

and $(q, \Delta t) = (0.65, 20)$ for the FC layer.

*4.5 Hardware and software*